\documentclass[aoas,preprint]{imsart}

\usepackage{amsthm,amsmath,amsfonts}
\usepackage{natbib}
\usepackage[colorlinks,citecolor=blue,urlcolor=blue,breaklinks]{hyperref}

\usepackage{graphicx}
\usepackage{bm}

\startlocaldefs
\DeclareMathSymbol{\Real}{\mathbin}{AMSb}{"52}
\def\pr{{\rm Pr}}
\def\v{{\varepsilon}}
\def\E{{\rm E}}
\def\calP{{\mathcal{P}}}
\endlocaldefs

\begin{document}

\begin{frontmatter}

\title{Bayesian inference for the Brown--Resnick process, with an application to extreme low temperatures}
\runtitle{Bayesian inference for the Brown--Resnick process}

\begin{aug}
\author{\fnms{Emeric} \snm{Thibaud}\corref{}\thanksref{m1}\ead[label=e1]{emeric.thibaud@colostate.edu}},
\author{\fnms{Juha} \snm{Aalto}\thanksref{m2}\ead[label=e2]{juha.aalto@fmi.fi}},
\author{\fnms{Daniel~S.} \snm{Cooley}\thanksref{m1}\ead[label=e3]{cooleyd@stat.colostate.edu}},
\author{\fnms{Anthony~C.} \snm{Davison}\thanksref{m3}\ead[label=e4]{anthony.davison@epfl.ch}}
\and
\author{\fnms{Juha} \snm{Heikkinen}\thanksref{m4}\ead[label=e5]{juha.heikkinen@luke.fi}}

\runauthor{E. Thibaud et al.}

\affiliation{Colorado State University\thanksmark{m1}, Finnish Meteorological Institute\thanksmark{m2}, Ecole Polytechnique F\'ed\'erale de Lausanne\thanksmark{m3} and Natural Resources Institute Finland\thanksmark{m4}}

\address{Address of the First and Third authors\\
Colorado State University\\
Department of Statistics\\
Fort Collins, CO 80523-1877, U.S.A.\\
\printead{e1}\\
\phantom{E-mail:\ }\printead*{e3}}

\address{Address of the Second author\\
Finnish Meteorological Institute\\
P.O Box 503\\
00101 Helsinki, Finland\\
\printead{e2}}

\address{Address of the Fourth author\\
Ecole Polytechnique F\'ed\'erale de Lausanne\\
EPFL-FSB-MATHAA-STAT, Station 8\\
1015 Lausanne, Switzerland\\
\printead{e4}}

\address{Address of the Fifth author\\
Natural Resources Institute Finland\\
P.O. Box 18\\
01370 Vantaa, Finland\\
\printead{e5}}

\end{aug}

\begin{abstract}
The Brown--Resnick max-stable process has proven to be well-suited for modeling extremes of complex environmental processes, but in many applications its likelihood function is intractable and inference must be based on a composite likelihood, thereby preventing the use of classical Bayesian techniques. In this paper we exploit a case in which the full likelihood of a Brown--Resnick process can be calculated, using componentwise maxima and their partitions in terms of individual events, and we propose two new approaches to inference. The first estimates the partitions using declustering,  while the second uses random partitions in a Markov chain Monte Carlo algorithm. We use these approaches to construct a Bayesian hierarchical model for extreme low temperatures in northern Fennoscandia.
\end{abstract}

\begin{keyword}
\kwd{Global warming}
\kwd{Likelihood-based inference}
\kwd{Max-stable process}
\kwd{Non-stationary extremes}
\kwd{Partition}
\kwd{Space-time declustering}
\end{keyword}

\end{frontmatter}

\section{Introduction}\label{sec:intro}

Extreme events such as heat or cold waves have major impacts on human health and ecological sustainability. Motivated by the pressing need for better risk assessment, inference for spatial extremes has developed rapidly during the last decade \citep[e.g.,][]{deHaan.Ferreira:2006,Davison.etal:2012,Davison.Huser.Thibaud:2013,Ribatet:2013}. Max-stable processes are the only possible non-degenerate limits for rescaled pointwise maxima of random processes, thus motivating their use in modeling spatial extremes.

Flexible max-stable models have been developed for inference on spatial extremes. Over a variety of applications, the Brown--Resnick process~\citep{Brown.Resnick:1977,Kabluchko.etal:2009} has proven to be a valuable model for extremes of environmental variables, such as rainfall, temperature, wind speed and river discharge \citep[see, e.g.,][]{Davison.Huser.Thibaud:2013,Engelke.etal:2012,Asadi.etal:2015,Buhl.Kluppelberg:2016}. It is built using Gaussian processes and its dependence structure is determined by a variogram, so  well-established ideas from classical geostatistics may be exported to the extremal context. The greater challenge is inference, since the full likelihood of the Brown--Resnick process is tractable only in relatively low-dimensional settings \citep{Huser.Davison:2013,Castruccio.etal:2015}. Composite likelihood techniques have been successfully used to fit max-stable models to block maxima or threshold exceedances, but with reduced estimation efficiency. More efficient procedures based on full, possibly censored, likelihoods \citep{Engelke.etal:2012,Wadsworth.Tawn:2013,Thibaud.Opitz:2015} have been developed for threshold exceedances, allowing the construction of more elaborate models for spatial extremes. 

Bayesian hierarchical models are appealing in spatial statistics because of the flexibility with which they can track the spatial variation of marginal distributions. \citet{Casson.Coles:1999}, \citet{Cooley.etal:2007} and \citet{Sang.Gelfand:2009} constructed hierarchical models for extremes assuming conditional independence of extreme occurrences. Such models may have flexible margins but cannot produce realistic simulations of extreme events \citep{Davison.etal:2012}.  \citet{Sang.Gelfand:2010} and \citet{Fuentes.etal:2013} used hierarchical models that model spatial dependence using Gaussian processes, a somewhat restrictive class of models that are asymptotically independent and thus may underestimate potential dependence at the most extreme levels \citep{Davison.Huser.Thibaud:2013}.
\citet{Ribatet.etal:2012} and \citet{Shaby:2013} proposed the use of composite likelihoods in Markov chain Monte Carlo (MCMC) algorithms and applied them for Bayesian inference on max-stable models, but these methods present computational challenges and do not yield exact simulation from the posterior distribution. \citet{Shaby.Reich:2012} constructed a max-stable spatial model based on a hierarchical representation of logistic extreme-value distributions \citep{Stephenson:2009}.

In this paper we propose a Bayesian hierarchical model based on the Brown--Resnick process, which exploits a special case for which the full likelihood of such a process can be calculated, using the joint distribution of componentwise maxima and their partitions in terms of individual events \citep{Stephenson.Tawn:2005}. We discuss the difficulties of using the partitions in practice, and we propose a new Bayesian approach to fitting max-stable processes without fixed partitions.

Our work is motivated by the changing climate of northern Fennoscandia. In northern regions of Norway, Finland, Russia and Sweden, extreme low temperatures play a central role in ecological sustainability by limiting the extent of outbreaks of forest pests such as the autumnal moth \emph{Epirrita autumnata},\ whose larvae have caused serious damage by defoliating mountain birch. Forests in the region have mostly been protected from this pest by extreme low winter temperatures, since its eggs cannot survive below $-36^\circ$C \citep{Virtanen.etal:1998}. Winter minimum temperatures are often below this threshold, but in recent years they have increased and projections suggest they will increase even more in the future \citep[][\S E.1]{IPCC:2013}. Although temperature is not the only factor limiting the spread of the moth, it is one of the most important, and thus plays a central role in the ecology of northern Fennoscandia. In \S\ref{sec:forecasts} we discuss the possible changes in winter minimum temperatures in the region, focusing on the probability that they will be less than $-36^\circ$C.

Motivated by recent disastrous heat wave events over the world, max-stable processes and related methods have been used to model temperature extremes and their potential changes in time. \citet{Davison.Gholamrezaee:2012} used the so-called \citet{Schlather:2002} process to model annual maximum temperatures in Switzerland. They used latitude, longitude and elevation to estimate response surfaces for the marginal parameters of extremes. \citet{Fuentes.etal:2013} describe a Bayesian spatial model for annual maximum temperatures in the US based on mixtures of Gaussian distributions. \citet{Shaby.Reich:2012} used a Bayesian max-stable model to model extreme temperature data over Europe. 

Section~\ref{sec:setting} describes the data we study. Section~\ref{sec:EVT} summarizes some extreme value ideas and introduces the inference methods for the Brown--Resnick model. The hierarchical model and the application are discussed in Section~\ref{sec:BHmodel}. 

\section{Extreme temperatures in northern Fennoscandia}\label{sec:setting}

\subsection{Dataset}\label{sec:dataset}

\begin{figure}[t]
\centerline{\includegraphics[width=0.81\textwidth]{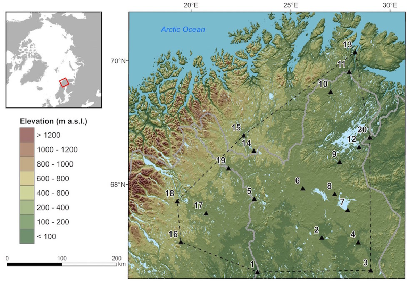}}
\caption{Locations of the $20$ weather stations in the northern regions of Finland, Norway and Sweden. The dashed lines shows the convex hull of the stations, which will be our region of interest for simulation.}
\label{fig:mapFinland}
\end{figure}

We use temperature data from 20 meteorological stations located in northern regions of Finland, Norway and Sweden; see Fig.~\ref{fig:mapFinland}. The smallest and largest distances between two stations are 26 and 471~km. Daily minimum temperatures are available from 1960 to 2013, but with data missing. In view of the motivation for our application, inference is needed only for the winter minima. We focus on modeling minimum temperatures in the months December--March and we use the daily data to investigate their co-occurrences. 

\begin{figure}[t]
\centerline{\includegraphics[]{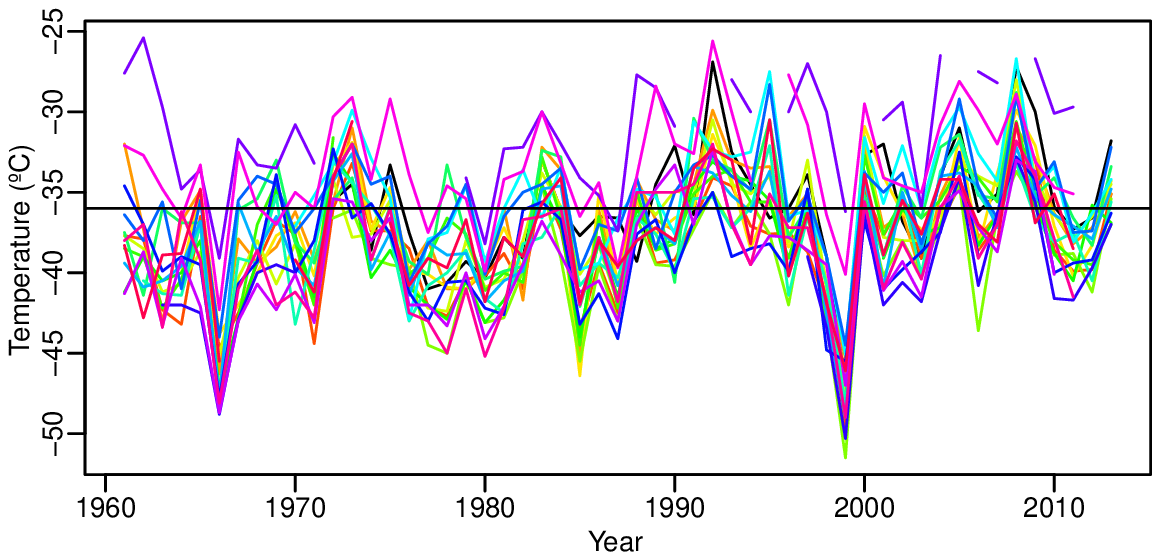}}
\caption{Winter minimum temperature ($^\circ$C) at the $20$ stations (a different color for each station). The horizontal line shows the $-36^\circ$C threshold.}
\label{fig:annualmin}
\end{figure}

Missing daily data censor the winter minima. Winter minimum values for the 20 stations are shown in Fig.~\ref{fig:annualmin}. Our dataset has at least 26 winter minima for each station, and each year has values of winter minimum for at least 15 of the 20 stations.

Mean winter minimum temperatures depend on local geography and topography, and we shall model this non-stationarity in their marginal distributions using covariates and random effects, which will also allow us to extrapolate results to ungauged sites.

We derived covariates that are known to influence minimum temperature in northern Fennoscandia \citep{Aalto.etal:2014}. We supplemented the coordinates of the stations with their absolute elevation above sea level and elevation relative to the surrounding areas (calculated on a 2~km~$\times$~2~km spatial moving average), a proximity index to the Arctic Ocean, calculated with a 160~km~$\times$~160~km moving window assuming that the effect of the ocean reaches 80~km inland, and a lake cover index, which expresses the proportion of area covered by lakes in a 3~km~$\times$~3~km window \citep[see][]{Aalto.etal:2014}.

\subsection{Temporal stationarity}

In Fig.~\ref{fig:annualmin} it is difficult to see the presence of a temporal trend with only 53 years of noisy data, but the figure suggests that minima may generally be slightly less extreme from 1990 onwards. The inclusion of a temporal effect in the model will help to assess this. 

\subsection{Serial dependence}\label{sec:serialdep}

Since we focus on modeling the winter minima, temporal dependence in the daily temperatures is not of interest in itself. But our approach requires us to partition minima in terms of independent generating events, which is complicated by the presence of temporal dependence in the daily data; see~\S\ref{sec:declustering}. Cold waves stay over the region for several days, so minimum temperatures tend to be very dependent from day to day. Preliminary analysis \citep[see the Supplementary Material,][]{SuppMat} showed that the dependence is strong for the first few days and vanishes after approximately five to ten days at each location.

\subsection{Spatial dependence}\label{sec:spatialdep}

We expect winter minimum temperatures to be dependent over large regions, and this is  the case in our data. Fig.~\ref{fig:annualmin} suggests that the minima for the $20$ stations behave similarly for the $53$ years: the years 1966 and 1999 were very cold at all $20$ locations, whereas 1973, 1993 and 2007 are warmer at all locations. It appears that most of the minima occurred on the same day or only a few days apart, so they may correspond to the same event, which suggests strong spatial dependence.

Results from a preliminary analysis \citep[see][]{SuppMat} suggest that pairwise distributions are asymptotically dependent and max-stable. After accounting for differences in the marginal distributions, we will use a stationary isotropic max-stable model for modeling spatial dependence.

\section{Max-stable processes and inference for spatial extremes}\label{sec:EVT}

\subsection{Generalities}

Let $Y=\{Y(\bm s)\}_{\bm s\in S}$ be a random process with continuous sample paths defined over a compact subset $S$ of $\Real^m$, and let $Y_1,Y_2,\ldots$ be independent replicates of $Y$. If there exist sequences of continuous functions $\{a_n(\bm s)\}_{\bm s\in S}>0$ and $\{b_n(\bm s)\}_{\bm s\in S}$ such that we have the distributional convergence
\begin{equation*}
\left\{{a_n(\bm s)}^{-1}\left[\max\{ Y_1(\bm s),\ldots,Y_n(\bm s) \} -b_n(\bm s)\right]\right\}_{\bm s\in S} \rightarrow \{Z(\bm s)\}_{\bm s\in S},\quad n\rightarrow\infty,
\end{equation*} 
where $Z=\{Z(\bm s)\}_{\bm s\in S}$ is a non-degenerate random process, we say that $Y$ lies in the max-domain of attraction of the process $Z$. Then the process $Z$ must be max-stable \citep{deHaan.Ferreira:2006} and its marginals are of generalized extreme-value (GEV) form,
\begin{equation*}
\pr\{ Z(\bm s) \leq z \} = \exp\left( -\left[ 1+ \xi(\bm s)\sigma(\bm s)^{-1} \{z-\mu(\bm s)\} \right]_+^{-1/\xi(\bm s)}\right), \quad \bm s\in S,
\end{equation*}
denoted $Z(\bm s)\sim{\rm GEV}\{\mu(\bm s),\sigma(\bm s),\xi(\bm s)\}$, with $x_+=\max(0,x)$, and real location $\mu(\bm s)$, scale $\sigma(\bm s)>0$ and shape $\xi(\bm s)$ parameters; throughout the paper, the case $\xi(\bm s)=0$ is defined as the limit for $\xi(\bm s)\rightarrow0$. Every max-stable process with marginal parameters $\mu(\bm s)$, $\sigma(\bm s)$ and $\xi(\bm s)$ can be written as $\mu(\bm s)+\sigma(\bm s)\xi(\bm s)^{-1}\{ Z(\bm s)^{\xi(\bm s)} -1 \}$ where $Z$ is a simple max-stable process with unit Fr\'echet margins, i.e., $\mu(\bm s)=\sigma(\bm s)=\xi(\bm s)=1$. Thus it suffices to consider simple max-stable processes, henceforth denoted by $Z$.

As a consequence of max-stability, for every set of locations ${\bm s_1,\ldots,\bm s_D\in S}$,
\begin{multline}\label{eq:distribMS}
\pr\{ Z(\bm s_1) \leq z_1,\ldots, Z(\bm s_D) \leq z_D \} = \exp\{ -V(z_1,\ldots,z_D) \},\\ \quad z_1,\ldots,z_D>0,
\end{multline}
where the so-called exponent function $V$ is homogeneous of order $-1$, i.e., $V(tz_1,\ldots,tz_D)=t^{-1}V(z_1,\ldots,z_D)$, $t>0$, and satisfies $V(z,\infty,\ldots,\infty)=z^{-1}$ for any permutation of its arguments. In particular, 
\begin{equation*}
\pr\{ Z(\bm s_1) \leq z,\ldots, Z(\bm s_D) \leq z \} = \exp(-\theta/z), \quad z>0,
\end{equation*}
where $\theta=V(1,\ldots,1)\in[1,D]$ is known as the extremal coefficient of $Z$ relative to $\bm s_1,\ldots,\bm s_D$. The extremal coefficient can be interpreted as the effective number of independent variables among $Z(\bm s_1),\ldots, Z(\bm s_D)$ and is useful for model checking, since it can be estimated non-parametrically.

As they appear as limits for rescaled pointwise maxima of random processes, max-stable processes are natural models for observed block maxima. Moreover, the equality $\min\{Y_1(\bm s),\ldots,Y_n(\bm s)\} = -\max\{-Y_1(\bm s),\ldots,-Y_n(\bm s)\}$ shows that such processes also provide natural models for block minima.

Several parametric max-stable models have been proposed. The \citet{Smith:1990} process was the first to be considered but its realizations are too smooth for realistic modeling of complex extremes. The \citet{Schlather:2002} process is constructed from stationary Gaussian processes and provides more realistic realizations controlled by different correlation functions. The extremal-$t$ process extends the Schlather process by introducing an additional shape parameter \citep{Opitz:2013}. Here we focus on the Brown--Resnick process \citep{Brown.Resnick:1977,Kabluchko.etal:2009} which appears to be a good model for extremes of environmental processes.

\subsection{Brown--Resnick process}

\hspace*{0cm}\citet{Kabluchko.etal:2009} constructed a stationary max-stable process, called the Brown--Resnick process, as
\begin{equation}\label{eq:defBR}
Z(\bm s) = \max_{i=1,2,\ldots} Q_i^{-1} W_i(\bm s), \quad W_i(\bm s)=\exp\{\v_i(\bm s) - \sigma(\bm s)^2/2 \}, \quad \bm s\in S,
\end{equation}
where the $Q_i$'s are the points of a unit rate Poisson process on $(0,\infty)$ and the $\v_i$ are independent replicates of an intrinsically stationary centered Gaussian process $\v$ with variance function $\sigma(\bm s)^2=\E\{\v(\bm s)^2\}$. The dependence structure of $Z$ is inherited from that of $\v$, which is controlled by the variogram $2\gamma(\bm h) = \E[\{ \v(\bm s+\bm h) - \v(\bm s) \}^2]$. Without loss of generality, we suppose that $\v(\bm 0)=0$ with probability one, so $\sigma(\bm s)^2=2\gamma(\bm s)$. A large class of models may be obtained by choosing different variograms. Later we use the stable variogram $2\gamma(\bm h) = 2(\|\bm h\|/\lambda)^{\kappa}$, with $\lambda>0$ and $\kappa \in (0,2]$. 

The exponent function of a Brown--Resnick process is \citep{Nikoloulopoulos.etal:2009}
\begin{multline}\label{eq:VBR}
V(z_1,\ldots,z_D) = \\ \sum_{j=1}^D z_j^{-1} \Phi_{D-1}\left\{ \log(\bm z_{-j}/z_j) ;  {\bm \Gamma}_{-j,j},{\bm \Gamma}_{-j,j}{\bm 1_{D-1}}'+{\bm \Gamma}_{j,-j}{\bm 1_{D-1}} - {\bm \Gamma}_{-j,-j} \right\},
\end{multline}
where $\Phi_p(\cdot; {\bm \mu},{\bm \Omega})$ is the cumulative distribution function of a $p$-dimensional Gaussian distribution with mean vector $\bm \mu$ and covariance matrix $\bm \Omega$, $\bm 1_{D-1}=(1,\ldots,1)' \in\Real^{D-1}$, $\bm \Gamma=\{\gamma(\bm s_{j_1}-\bm s_{j_2})\}_{1\leq j_1,j_2\leq D}$, ${\bm \Gamma}_{-j,j}$ denotes the $(D-1)\times 1$ vector consisting of the $j$th column of ${\bm \Gamma}$ without its $j$th component,  ${\bm \Gamma}_{-j,-j}$ denotes the matrix ${\bm \Gamma}$ without its $j$th row and $j$th column, and so forth.

Let $\bm Z=\{Z(\bm s_1),\ldots,Z(\bm s_D)\}$ be a vector stemming from a simple max-stable process $Z$. Differentiation of~\eqref{eq:distribMS} shows that the full likelihood contribution for one realization $(z_1,\ldots,z_D)$ of $\bm Z$ may be written as 
\begin{equation}\label{eq:likMS}
\sum_{\pi=(\pi_1,\ldots,\pi_K) \in \calP} \left[\prod_{k=1}^K \left\{-V_{\pi_k}(z_1,\ldots,z_D)\right\} \right] \exp\{-V(z_1,\ldots,z_D)\},
\end{equation}
where $\calP$ denotes the set of all partitions of $\{1,\ldots,D\}$ and the subscript $\pi_k$ on $V_{\pi_k}$ denotes the partial derivative of the function $V$ with respect to the components in the subset $\pi_k$. Although the likelihood has a closed-form expression for the Brown--Resnick process \citep{Wadsworth.Tawn:2013}, the size of $\calP$ equals the Bell number and is larger than $10^5$ for $D\geq 10$, making full likelihood inference computationally unattainable in many situations. Hence inference for max-stable processes has typically  been based on composite likelihoods using lower-order densities \citep{Padoan.etal:2010,Castruccio.etal:2015}. 
Not only do composite likelihoods result in a loss of efficiency compared to full likelihood estimation, they also increase the computational burden when employed in a Bayesian analysis, because the composite likelihood does not directly yield a posterior which appropriately accounts for parameter uncertainty \citep[][]{Ribatet.etal:2012,Shaby:2013}. \citet{Engelke.etal:2012}, \citet{Wadsworth.Tawn:2013} and \citet{Thibaud.Opitz:2015} developed full likelihood inferences for threshold exceedances of processes in the max-domain of attraction of a Brown--Resnick process, which, while challenging, have lower computational burdens than does~\eqref{eq:likMS}.

For our likelihood, we employ the result of \citet{Wadsworth.Tawn:2013}, who showed that, for the Brown--Resnick process, partial derivatives of $V$ with respect to the first $d$ indices, denoted $V_{1:d}$, satisfy
\begin{multline}\label{eq:derivV}
-V_{1:d}(\bm z)= \dfrac{\Phi_{D-d}\left( \log\bm z_{(d+1):D};\tilde{\bm\tau}, \tilde{\bm\Gamma} \right)}{(2\pi)^{(d-1)/2} |\bm\Sigma_{1:d}|^{1/2}(\bm 1_d' \bm q_d)^{1/2}\prod_{j=1}^d z_j}\\
\\ \times \exp\left\{ -\dfrac{1}{2}\left( \dfrac{1}{4}\bm\sigma_d' \bm\Sigma_{1:d}^{-1} \bm\sigma_d - \dfrac{1}{4} \dfrac{\bm\sigma_d' \bm q_d \bm q_d' \bm\sigma_d}{\bm 1_d' \bm q_d} +\dfrac{\bm\sigma_d' \bm q_d}{\bm 1_d' \bm q_d} - \dfrac{1}{\bm 1_d' \bm q_d} \right) \right\}\\
\\ \times \exp\left[ -\dfrac{1}{2}\left\{ \log\bm z_{1:d}' \bm A_{1:d} \log\bm z_{1:d} + \log\bm z_{1:d}' \left(\dfrac{2\bm q_d}{\bm 1_d' \bm q_d}\right) + \bm\Sigma_{1:d}^{-1} \bm\sigma_d - \dfrac{\bm q_d \bm q_d' \bm\sigma_d}{\bm 1_d' \bm q_d} \right\} \right],
\end{multline}
with $\tilde{\bm\Gamma} = (\bm K_{01}' \bm A \bm K_{01})^{-1}$, $\tilde{\bm\tau} = -\tilde{\bm\Gamma}(\bm K_{01}' \bm A \bm K_{01} \log \bm z_{1:d} + \bm K_{01}'\bm\Sigma^{-1}\bm 1_D/\bm 1_D'\bm\Sigma^{-1}\bm 1_D)$, $\bm\Sigma=\{\gamma(\bm s_{j_1})+\gamma(\bm s_{j_2})-\gamma(\bm s_{j_1}-\bm s_{j_2})\}_{1\leq j_1,j_2\leq D}$, $\bm\Sigma_{1:d}$ the $d\times d$ matrix consisting of the first $d$ rows and columns of $\bm\Sigma$, $\bm A=\bm\Sigma^{-1}-\bm q\bm q'/ \bm 1_D' \bm q$, $\bm A_{1:d}=\bm\Sigma_{1:d}^{-1}-\bm q_d\bm q'_d/ \bm 1_d' \bm q_d$, $\bm q = \bm\Sigma^{-1} \bm 1_D$, $\bm q_d = \bm\Sigma_{1:d}^{-1} \bm 1_d$, $\bm\sigma={\rm diag}(\bm\Sigma)$ and $\bm\sigma_d=\bm \sigma_{1:d}$, where 
$$
\bm K_{10}=
\left(
\begin{array}{c}
\bm I_d\\
\bm 0_{D-d,d}   
\end{array}
\right)
,\quad \bm K_{01}=
\left(
\begin{array}{c}
\bm 0_{d,D-d}\\
\bm I_{D-d}
\end{array}
\right).
$$

\subsection{The Stephenson--Tawn model}\label{sec:STmodel}

\hspace*{0cm}\citet{Stephenson.Tawn:2005} derived a likelihood based on componentwise block maxima and their occurrence times that avoids direct differentiation of~\eqref{eq:distribMS}. Let $\bm Y_1, \bm Y_2,\dots$ denote independent replicates of a random vector $\bm Y=\{Y(\bm s_1),\ldots,Y(\bm s_D)\}\in\Real^D$. Without loss of generality, suppose that the process $Y$ has unit Fr\'echet margins and lies in the max-domain of attraction of a simple max-stable process $Z$. Let $\bm M_n = \max_{i=1,\ldots,n} \bm Y_i/n$ denote the componentwise maxima, and let $\Pi_n=(\pi_1,\ldots,\pi_K)$ denote the unique partition of $\{1,\ldots,D\}$ such that $\{M_{n,j}\}_{j\in \pi_k}$ comes from the same individual observation, i.e., for each $k\in\{1,\ldots, K\}$, there exists $i_k$ such that $\{M_{n,j}\}_{j\in \pi_k}=\{Y_{i_k,j}/n\}_{j\in \pi_k}$. Then \citet{Stephenson.Tawn:2005} showed that, as $n\rightarrow\infty$, the block maximum and its associated partition $(\bm M_n,\Pi_n)$ converge in distribution to $(\bm Z,\Pi)$, whose joint density is
\begin{equation}\label{eq:likST}
f(\bm Z,\Pi) = \exp\{-V(\bm Z)\} \prod_{\pi_k \in \Pi} \left\{-V_{\pi_k}(\bm Z) \right\}.
\end{equation}
This suggests using~\eqref{eq:likST} as a likelihood for inference on the joint distribution of block maxima and their occurrence times.

This approach only requires the maximum and the partition to be observed, and thus represents a compromise between the block maximum and threshold exceedance approaches. The relative efficiency of Stephenson--Tawn and pairwise maximum likelihood estimators has been investigated by \citet{Wadsworth.Tawn:2013}. Because the partition contains information about the dependence, the Stephenson--Tawn estimator has smaller variance than that based only on the block maxima. However, for finite block length, the former may be more biased, because the model $(\bm Z,\Pi)$ is misspecified for $(\bm M_n,\Pi_n)$, i.e., both for the maximum and the partition \citep{Wadsworth:2015}. Use of the limiting distribution for $(\bm M_n,\Pi_n)$ represents a strong assumption that can lead to poor fit, whereas approaches based on the maximum $\bm M_n$ alone are more robust.

A further difficulty arises from the definition of the partition for serially dependent data. The partition in the Stephenson--Tawn model is defined in terms of independent generating events. While it is natural to define the partition in terms of the days of occurrence of the maxima when the daily observations are independent, its estimation is harder when data show temporal dependence. In \S\ref{sec:declustering} we propose a space-time declustering procedure to estimate the partition. 

\subsection{Random partition model}\label{sec:estpart}

Although the use of the estimated fixed partition is theoretically appealing, it represents a practical challenge. First, the model is misspecified, which affects the estimation of the max-stable model by the introduction of potentially biased information. Second, the partition must be estimated in a preliminary step, and  inference based on the Stephenson--Tawn model is conditional on the fixed partition and ignores the uncertainty associated to its estimation. The partition is not always very well defined (see \S\ref{sec:declustering}) and its estimation is sensitive to measurement error. This motivates us to propose an approach to estimating the parameters of the Brown--Resnick process, without fixing the partition, which uses only the values of the maxima and can be applied when the occurrence dates are unavailable.

Let $\bm Z=\{Z(\bm s_1),\ldots,Z(\bm s_D)\}$ be a vector stemming from a Brown--Resnick process $Z$. To overcome the intractability of the likelihood~\eqref{eq:likMS}, we propose to augment the data with the unknown partition $\Pi$ that generates $\bm Z$ in terms of the functions $P_i=Q_i^{-1}W_i(\bm s)$ in~\eqref{eq:defBR}, i.e., $\Pi=(\pi_1,\ldots,\pi_K)$ is defined such that the points $Z(\bm s_j)$ in the same set $\pi_k$ come from the same function $P_i$. The joint distribution of $(\bm Z,\Pi)$ is~\eqref{eq:likST}.

We impute the unknown partition $\Pi$ given the observed value of $\bm Z$ in an MCMC algorithm, using ideas similar to \citet{Dombry.etal:2013}, who considered the partition in the context of conditional simulation from max-stable processes. The partition is imputed from its conditional distribution given the data. From~\eqref{eq:likST} we have
\begin{equation}\label{eq:partdistr}
\Pr(\Pi=\pi \mid \bm Z=\bm z) = \dfrac{f(\bm z,\pi)}{\sum_{\tilde \pi\in\calP} f(\bm z,\tilde\pi)}.
\end{equation}
However, the number of partitions of the set $\{1,\ldots,D\}$ is the $D$-th Bell number, so~\eqref{eq:partdistr} is intractable for large $D$. We use a Gibbs sampler to update the partition. For a partition $\pi=(\pi_1,\ldots,\pi_K)$, we choose a location $\bm s_j$ at random and let $\pi_{-j}$ denote the restriction of $\pi$ to $\{\bm s_1,\ldots,\bm s_D\}\setminus\{\bm s_j\}$. The size of $\pi_{-j}$ is $K$, or $(K-1)$ if $\{\bm s_j\}$ is itself a partitioning set of $\pi$. Given $\pi_{-j}$, the partition is updated by attributing the location $\bm s_j$ either to an existing set of $\pi_{-j}$ or to a new set, giving at most $K+1$ choices. The conditional probability for a new partition $\pi^*$ with $\pi^*_{-j}=\pi_{-j}$ is
\begin{equation}\label{eq:partconddist}
\Pr(\Pi=\pi^* \mid \Pi_{-j}=\pi_{-j}, \bm Z=\bm z) = \dfrac{f(\bm z,\pi^*)}{\sum_{\tilde\pi\in\calP: \tilde\pi_{-j}=\pi_{-j}} f(\bm z,\tilde\pi)}.
\end{equation}
The discrete distribution~\eqref{eq:partconddist} has support on a set of size at most $K+1$, so a new partition $\pi^*$ can easily be simulated.

In practice, inference is based on independent replicates $\bm Z_1,\ldots, \bm Z_N$, and the MCMC algorithm independently updates the corresponding latent partitions $\Pi_1,\ldots, \Pi_N$. Given the observed values of the $\bm Z_i$ and the simulated $\Pi_i$, the marginal and dependence parameters of the model are updated from the joint distribution of the $(\bm Z_i,\Pi_i)$, which is a product of terms of the form~\eqref{eq:likST}.

\section{Modeling extreme temperatures in northern Fennoscandia}\label{sec:BHmodel}

\subsection{Exploratory analysis}\label{sec:prelim}

We first fitted GEV distributions separately at each of the 20 stations; see \citet{SuppMat}. The estimates for the scale and shape parameters were not significantly different from site to site, but variation in the location parameter estimates suggested use of a non-stationary GEV model with spatially varying location parameter and constant scale and shape parameters.

We then fitted non-stationary marginal models at the 20 locations by maximizing the independence likelihood, which assumes the independence of the temperatures from site to site. Non-stationarity was modeled using trend surfaces defined from the covariates mentioned in~\S\ref{sec:dataset} and temporal trends in the location and scale parameters.  We used the R package SpatialExtremes \citep{SpatialExtremes:2015} to fit these models, which were compared using the Takeuchi information criterion, an equivalent of the Akaike information criterion for misspecified models. Our best model had all six covariates for the location surface, constant scale and shape parameters, and a temporal trend in the location only. Its fit was assessed using marginal QQ-plots: departure from the $95\%$ confidence intervals was observed for five stations. This suggests that the trend surface model is not  flexible enough to model the variability in the location parameters over the 20 stations. Inclusion of additional covariates did not improve the fit, motivating the use of a more flexible approach using random effects in a Bayesian hierarchical model. 

\subsection{Hierarchical model}\label{sec:hiermodel}

\subsubsection{Definition}

We consider two Bayesian hierarchical models based on the Brown--Resnick process. The first uses the Stephenson--Tawn model and estimates the partitions in a preliminary step through space-time declustering; see~\S\S\ref{sec:STmodel},~\ref{sec:declustering}. The second imputes the partitions in the MCMC algorithm; see~\S\ref{sec:estpart}.  Here we describe the part of the hierarchical model for the data conditional on the partitions, which is common to both approaches.

Let $\bm y_i=(y_{i1},\ldots,y_{iD})$ ($i=1,\ldots,N$) denote the observed minimum temperatures for the $N$ years of data, let $\bm s_1,\ldots,\bm s_D$ denote the $D=20$ locations of the stations such that $y_{ij}=y_i(\bm s_j)$, let $t_{ij} \in\Real$ denote the covariate for the day of occurrence of the winter minimum $y_{ij}$, defined as the number of years from 1 January 2000, and finally let $\pi_i=(\pi_{i1},\ldots,\pi_{i K_i})$ denote the partitions of $\{1,\ldots,D\}$ for each year. When data are missing, the dimension of $\bm y_i$ is lower than $D$ and so is the size of the corresponding partition. The first level of the hierarchical model gives the probability density of the observed minima and the partitions $\pi_1,\ldots, \pi_N$, conditional on the marginal and dependence parameters $\bm\psi$, as
\begin{multline}\label{eq:likST2} 
L\left(\{\bm y_i\}, \{\pi_i\} \mid \bm\psi\right)= \\ \prod_{i=1}^N \left( \exp[-V\{f_{i1}(-y_{i1}),\ldots,f_{iD}(-y_{iD})\}] \prod_{k=1}^{K_i} \left[-V_{\pi_{ik}}\{f_{i1}(-y_{i1}),\ldots,f_{iD}(-y_{iD})\} \right] \prod_{j=1}^D f_{ij}'(-y_{ij})  \right) ,
\end{multline}
where 
\begin{equation*}
f_{ij}(y) = \left\{1+\xi_{ij}\sigma_{ij}^{-1}(y-\mu_{ij})\right\}_+^{1/\xi_{ij}}
\end{equation*}
denotes the marginal transformation to the standard Fr\'echet scale for $\xi_{ij}\neq0$ and the prime denotes derivative with respect to $y$; the negative signs for the $y_{ij}$'s in~\eqref{eq:likST2} correspond to modeling negated winter minima.

The second level of the hierarchical model specifies the models for the exponent measure $V$ and the GEV parameters. We define a space-time model for $\mu$ with the covariates discussed in~\S\ref{sec:dataset}, with a linear temporal trend, and with a Gaussian prior distribution as
\begin{equation*}
\mu(\bm s,t) \sim{\rm GP}\big\{ \bm X(\bm s) \bm\beta, \tau^2\rho \big\} + \alpha t
\end{equation*}
where $\bm X(\bm s)$ denotes the vector of the intercept and covariates at location $\bm s$, and ${\rm GP}\{x(\bm s),\tau^2\rho(\cdot)\}$ denotes a Gaussian process with mean $x(\bm s)$, variance $\tau^2$, and correlation function $\rho(\bm h)=\exp(-\|\bm h\|/\delta)$. The random effect  allows the location parameters to vary smoothly over space, in addition to any effect of the covariates. We kept $\sigma$ and $\xi$ constant over space and time because this appeared to suffice for a good fit. Thus $\mu_{ij}=\mu(\bm s_j,t_{ij})$, $\sigma_{ij}=\sigma$, and $\xi_{ij}=\xi$ for each pair of replicate and location indices $(i,j)$, and the marginal GEV distributions of the model at the observed data can be written as
\begin{equation*}
-y_{ij} \sim {\rm GEV}\left( U_j +\alpha t_{ij}, \sigma, \xi \right),
\end{equation*}
where $\bm U=(U_1,\ldots,U_D)$ is a Gaussian vector with mean $\bm X\bm\beta$, where $\bm X$ is a $D\times 7$ matrix with rows $\bm X(\bm s_j)$, and covariance matrix $\{ \tau^2 \rho(\bm s_{j_1}-\bm s_{j_2}) \}_{1\leq j_1,j_2 \leq D}$. We use a Brown--Resnick process with variogram $2\gamma(\bm h)=2(\|\bm h\|/\lambda)^\kappa$ as a model for the exponent function $V$.

\subsubsection{Prior distributions}

Prior distributions are assigned to the variables at the last level of a hierarchical model. We used vague priors for all parameters except those of the latent Gaussian field; see \citet{SuppMat}. The prior distributions for $\bm\beta$ and $\tau^2$ were chosen to obtain conjugate full conditional distributions for these parameters.

\subsubsection{Space-time declustering}\label{sec:declustering}

The use of the Stephenson--Tawn model involves choosing the partitions in~\eqref{eq:likST2}, whose sets should correspond to independent individual events. Several approaches have been proposed to identify clusters in univariate time series (e.g.,~\citealp[Chapter~10]{Beirlant.etal:2004}; \citealp{Chavez-Demoulin.Davison:2012}). In the multivariate case, declustering techniques have been proposed by \citet{Coles.Tawn:1991} and \citet{Nadarajah:2001}. Here we follow \citet{Coles.Tawn:1991}, who identified clusters of storm surges by assuming that events separated by a fixed time lag are independent. 

In our application, a difficulty arises from the use of the dates of occurrences. Because of the measurement precision of the temperatures ($0.1^\circ$C), the same annual minimum may occur on different days, affecting the estimation of the partition; this is the case for 7\% of the minima, with dates of occurrences separated by more than two days in 3\% of cases. When a minimum has multiple days of occurrences, we choose one of these days at random to define our partition.

For our temperature data, preliminary analysis suggests choosing a time lag of five days: two observations at the same location and separated by more than five days are assumed to be independent. To identify clusters for the 20 time series, we use a single linkage (or friend-to-friend) method whereby a cluster is composed of data that are dependent in pairs. Thus if the minima occurred at time 1 for the first series, time 4 for the second and time 7 for the third, these three minima are considered to lie in the same cluster. With this procedure, the size of the partition for each year is at most five, and winter temperature minima usually occurred from widespread events that lasted for between five and ten days.

\subsection{MCMC algorithm}
The MCMC algorithm for our Brown--Resnick model is nonstandard because the likelihood~\eqref{eq:likST2} cannot be calculated exactly but must be estimated. The likelihood involves the function $V$ and its partial derivatives, which themselves involve the Gaussian cumulative distribution function in their expressions, and thus can only be estimated. We used the approach of \citet{Andrieu.Roberts:2009} to use a likelihood estimator in the MCMC algorithm. Under the assumption that the likelihood estimator is unbiased, that algorithm generates a Markov chain that has the targeted posterior distribution as stationary distribution.

We estimated the likelihood~\eqref{eq:likST2} by replacing the Gaussian cumulative distribution functions in its expression by Monte Carlo estimators. Our estimator is not unbiased, but we constructed the estimator to minimize the potential bias in the estimation of the posterior and obtain good mixing properties for the Markov chain in a reasonable time; see \citet{SuppMat}. For our dataset and the fixed partitions estimated in~\S\ref{sec:declustering}, a single estimation of the likelihood~\eqref{eq:likST2} takes about one second on a single CPU.

We used a Gibbs sampler with Metropolis--Hastings steps to update the marginal and dependence parameters of the model. To minimize the number of likelihood estimations in the MCMC algorithm, and because some parameters were found to be very dependent, we updated the parameters of the model in blocks; see \citet{SuppMat}. With our choice of blocks, one iteration of the MCMC algorithm requires four likelihood estimations for the Stephenson--Tawn model. For the random partition model, the Gibbs sampler for the partitions requires estimating the probabilities~\eqref{eq:partconddist}, which increases the computational burden a bit more.

Since one iteration of the MCMC algorithm takes several seconds on a single CPU, we used parallel computing, fitting 50 independent Markov chains each of length 15,000 and discarding the first 5000 iterations of each chain. The resulting chains were merged to obtain a final chain of length 500,000. The MCMC diagnostics are discussed in \citet{SuppMat}. The mean of the posterior distributions were used as pointwise estimators. Credible intervals were calculated from the quantiles of the Markov chains.

The results of a simulation study suggest that the MCMC algorithm provides sensible inferences on the model parameters: the posterior means were close to the true values and the empirical coverages of the credible intervals were close to the nominal values; see \citet{SuppMat}.

\subsection{Results}

We compare three non-stationary Brown--Resnick models. The first, M1, uses the same trend surfaces as does the best marginal model discussed in~\S\ref{sec:prelim} and is fitted using pairwise likelihood as implemented in the R package SpatialExtremes; this is our baseline model, to which we compare our new approaches. The second, M2, is the Bayesian hierarchical model that uses the fixed partitions from the declustering procedure. The third, M3, is the Bayesian hierarchical model with the random partitions.

\begin{table}[t]
\caption{Estimated posterior mean parameters and $95\%$ confidence/credible intervals (in parentheses) from the trend surface model M1, the Bayesian model based on declustering M2 and the random partition Bayesian model M3. The second column is the estimated average location parameter $\mu_j=\mu(\bm s_j,0)$ over the 20 stations.}
\label{tab:estimates}
\begin{tabular}{lcccccc}
\hline
Model & $-\overline{\hat\mu_j}$ & $-\hat\alpha$ & $\hat\sigma$ & $\hat\xi$ & $\hat\lambda$ & $\hat\kappa$ \\
\hline
M1 & $-34.9$ & $0.07$ & $3.4$ & $-0.15$ & $371$ & $0.40$\\
 & \tiny{$(-35.8,-34.1)$} & \tiny{$(0.03,0.11)$} & \tiny{$(3.0,3.7)$} & \tiny{$(-0.19,-0.10)$} & \tiny{$(40,702)$} & \tiny{$(0.34,0.47)$}\\
M2 & $-34.6$ & $0.09$ & $4.0$ & $-0.11$ & $1811$ & $0.54$\\
 & \tiny{$(-35.6,-33.7)$} & \tiny{$(0.05,0.13)$} & \tiny{$(3.6,4.6)$} & \tiny{$(-0.15,-0.06)$} & \tiny{$(977,3615)$} & \tiny{$(0.46,0.63)$}\\
M3 & $-35.0$ & $0.06$ & $3.5$ & $-0.10$ & $1086$ & $0.53$\\
 & \tiny{$(-36.0,-33.8)$} & \tiny{$(0.00,0.11)$} & \tiny{$(3.0,4.1)$} & \tiny{$(-0.14,-0.06)$} & \tiny{$(466,3143)$} & \tiny{$(0.43,0.65)$}\\
\hline
\end{tabular}
\end{table}

\begin{figure}[t]
\centerline{\includegraphics[]{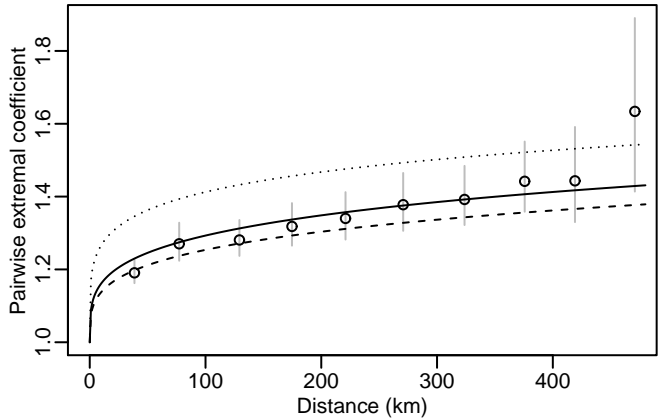}}
\caption{Binned estimates of empirical pairwise extremal coefficients (black, with their $95\%$ confidence intervals in gray). Curves are from the fitted Brown--Resnick models (see Table~\ref{tab:estimates}): M1 (dotted), M2 (dashed), and M3 (plain).}
\label{fig:extremalcoef}
\end{figure}

\begin{figure}[t]
\centerline{\includegraphics[width=1\textwidth]{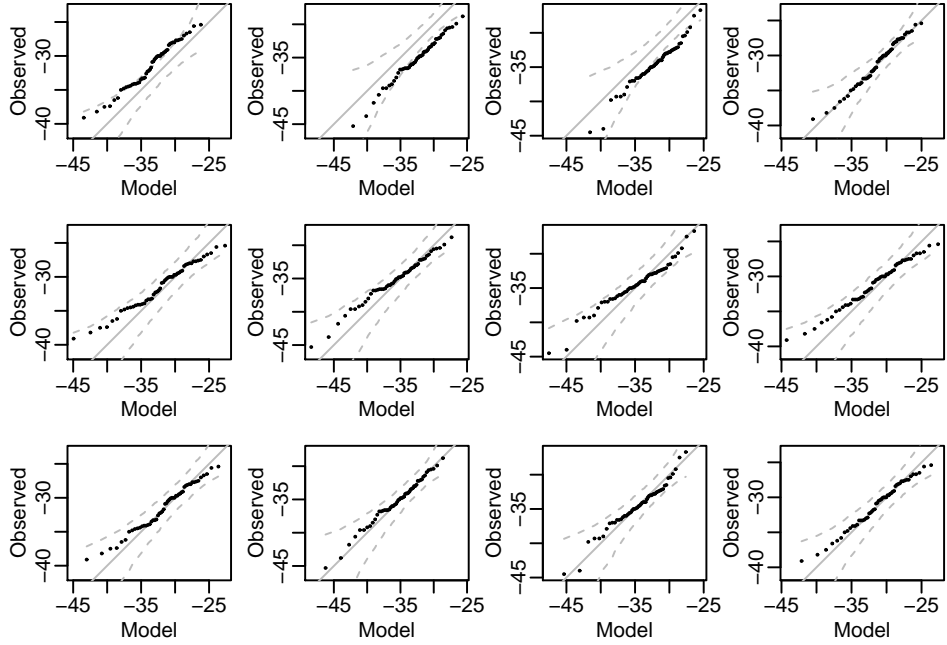}}
\caption{Model checking for the three Brown--Resnick models (see Table~\ref{tab:estimates}): M1 (top row), M2 (middle row) and M3 (bottom row). The QQ-plots compare the predictions and the observed maximum of winter minimum temperatures ($^\circ$C) over some groups of stations, $m_i=\max_{j\in J}y_{ij}$ ($i=1,\ldots,53$). The columns, from left to right, correspond to the group of stations $J=\{5,16,17,18,19\}$ (western region), $J=\{2,3,4,6,7,8,9,12,20\}$ (eastern region), $J=\{5,6,8,9,10,11,12,13,14,15,19,20\}$ (northern region), and all 20 stations. Grey dashed lines are $95\%$ overall confidence bands, and the solid grey line indicates a perfect fit.}
\label{fig:modelchecking_max}
\end{figure}

We discuss the estimates in terms of the negated location parameter $\mu$, and negated trend $\alpha$, which correspond to the original scale of minimum temperatures.  The estimates for the three models are given in Table~\ref{tab:estimates}. We assessed the fit of the marginal distributions using QQ-plots \citep[see][]{SuppMat}. We assessed the fit of the dependence model in terms of empirical pairwise extremal coefficients, computed using the empirical marginal distributions of the data to avoid potential bias due to the estimation of the marginals \citep{Ribatet:2013}; see Fig.~\ref{fig:extremalcoef}. As a global assessment of the fit in terms of the marginals and dependence, we computed the maximum of winter minimum temperatures over four different sets of stations, for the 53 years of the dataset, and compared the observed values with the predictions from the three Brown--Resnick models, computed by simulation using the algorithm of \citet{Dieker.Mikosch:2015}; see  Fig.~\ref{fig:modelchecking_max}. Further diagnostics for the prediction of the minimum and the averages of winter minimum temperatures are given in \citet{SuppMat}.

For the trend surface model, M1, the marginal QQ-plots show poor fits for six stations. Moreover, the model severely underestimates the dependence in terms of the pairwise extremal coefficient. This can be explained by the inadequacy of the fitted marginal distributions, which bias the dependence estimates; \citet{Ribatet:2013} also observed this when modeling wind gust data. In terms of predicting the maxima over groups of stations, the model performs poorly: many points lie outside the $95\%$ confidence bounds. In general, this model does not predict extreme temperatures well.

For the Bayesian model based on the declustering, M2, the marginal QQ-plots show that the sample quantiles are outside the $95\%$ confidence bands for three stations, and underestimate them  systematically at all stations; the estimated scale parameter $\sigma$ seems too large. The model overestimates the pairwise dependence somewhat, and tends to overestimate return levels for the maxima over different groups of stations.

For the random partition model, M3, the marginal QQ-plots show a good fit of the GEV distributions: the empirical quantiles are in the $95\%$ confidence bands for all the stations. The empirical pairwise extremal coefficients agree adequately with the curve from the model. In terms of the multivariate predictions, M3 can predict the maximum, minimum and average of the minima better than does M2, with all the observations within the $95\%$ confidence bounds; see Fig.~\ref{fig:modelchecking_max} and \citet{SuppMat}. This model appears to represent both the marginal and spatial behavior of extreme low temperatures adequately.

\begin{figure}[t]
\centerline{\includegraphics[width=1\textwidth]{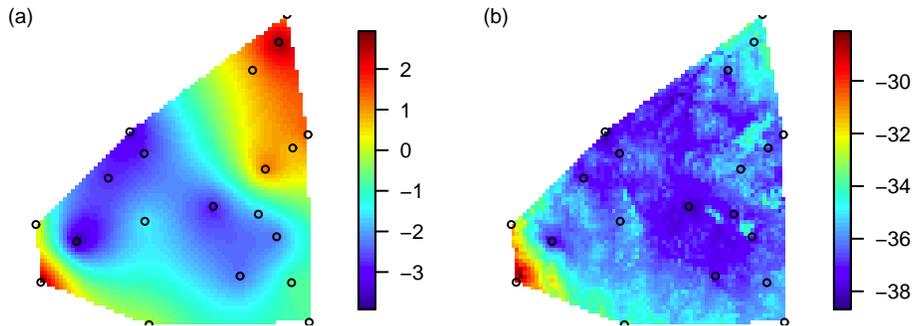}}
\caption{Posterior mean of the random effect in the location parameter (a) and mean of the GEV distribution for the year 2016 (b). The scale corresponds to temperature minima~($^\circ$C).}
\label{fig:mu_mean2016}
\end{figure}

We further discuss the results for the random partition Bayesian model M3. Fig.~\ref{fig:mu_mean2016}(a) shows the posterior mean of the random effects for the location parameter over the convex hull of the meteorological stations (see Fig.~\ref{fig:mapFinland}), and the mean of the GEV distribution for the year 2016. We restrict our analysis to the convex hull of the stations because extrapolation outside this region is too uncertain. The posterior mean of the random effect in the location parameters shows the variability in this parameter that cannot be explained by the covariates. The posterior mean for the fitted GEV distributions shows lower temperatures in the inland region.  The minima decrease as latitude, longitude, relative elevation and distance from the Arctic ocean increase, and increase as lake cover and elevation increase.  Similar covariate effects were found by \citet{Aalto.etal:2014} for extreme low temperatures in northern Fennoscandia. The estimated shape parameter, corresponding to Weibull---so-called light-tailed---distributions, agrees with previous studies on extreme temperatures. The dependence parameters correspond to strong dependence, with pairwise extremal coefficient $\hat\theta=1.42~(1.34,1.50)$ at 400~km, while $\hat\kappa=0.53~(0.43,0.65)$ implies that the realizations from the Brown--Resnick process are continuous but nowhere differentiable.

\subsection{Forecasts for extreme low temperatures for 2016--2030}\label{sec:forecasts}

\begin{figure}[t]
\centerline{\includegraphics[width=1\textwidth]{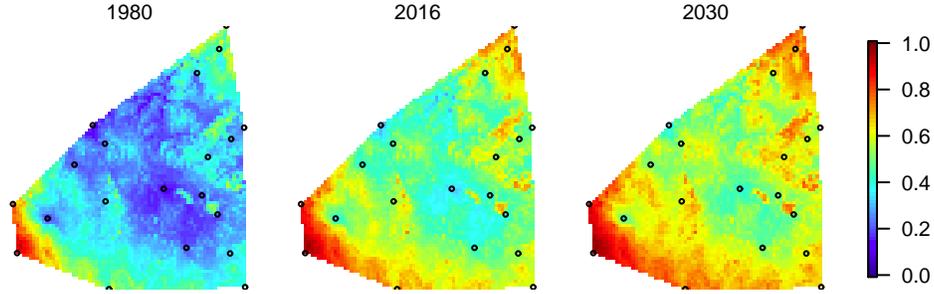}}
\caption{Probability that the winter minimum temperature is higher than $-36^\circ$C for 1980, 2016 and 2030.}
\label{fig:prob36}
\end{figure}

We base our forecasts on the random partition Bayesian model M3. Changes in extremes are entirely based on the linear trend $\alpha$ in the location parameter. This translates to changes in the mean of the GEV distribution at location $\bm s$ and year $t$, which is 
$$
\mu(\bm s,0) + \alpha t + \sigma\xi^{-1}\{\Gamma(1-\xi)-1\}, \quad \xi<1, \xi\neq 0.
$$
The estimate $-\hat\alpha=0.06~(0.00,0.11)$ corresponds to an increase of $0.6^\circ$C per decade in the mean winter minimum temperature. Fig.~\ref{fig:mu_mean2016}(b) shows the predicted mean winter minimum temperature for 2016. Fig.~\ref{fig:prob36} shows the probability that winter minimum temperatures are higher than $-36^\circ$C, the threshold needed for protecting northern Fennoscandia forests from outbreaks of \emph{Epirrita autumnata},\ for 1980, 2016 and 2030. There is a strong change in these probabilities. Over the region defined by the stations, the winter minimum temperatures had low probability of exceeding $-36^\circ$C before 1980, rising to about $0.54$ for 2016, and then to $0.61$ for 2030, according to the fitted model. The predictions must be interpreted with care, owing to the large uncertainty in the estimation of $\alpha$.

\begin{figure}[t]
\centerline{\includegraphics[width=1\textwidth]{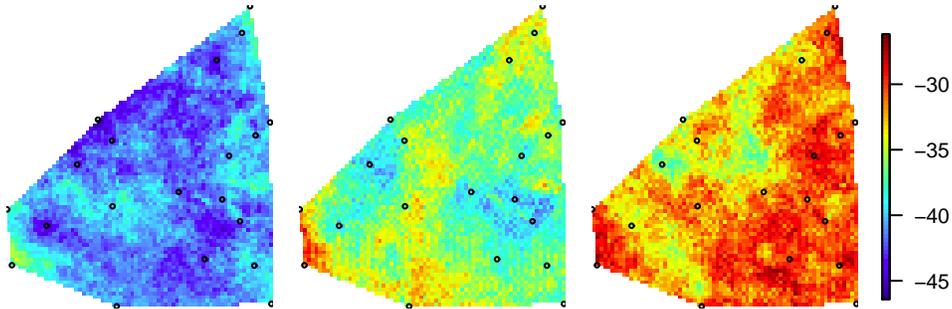}}
\caption{Three simulated minimum temperature fields ($^\circ$C) for 2016.}
\label{fig:simuBR}
\end{figure}

Spatial simulation of winter minimum temperatures is useful for estimating joint return levels over certain regions, which could be useful for assessing the risks of pest outbreaks over a region for example. Fig.~\ref{fig:simuBR} shows three simulations of winter minima fields from M3, which show how the strong dependence in the Brown--Resnick model results in winter minima that are either low or high throughout the region. The spatial heterogeneity is mostly due to non-stationarity in the marginal distributions.  Conditional simulations are possible \citep{Dombry.etal:2013} and could be useful as input to models to link past pest outbreaks with extreme temperatures, for example.

\section{Discussion}

Although the Stephenson--Tawn approach was found to perform well on simulated data when the true partitions are known, the results of our application show that it can be inappropriate in practice. The hierarchical model for the marginals does not give a good fit to the marginal GEV distributions, notwithstanding the additional flexibility provided by the random effects. 

\begin{figure}[t]
\centerline{\includegraphics[width=1\textwidth]{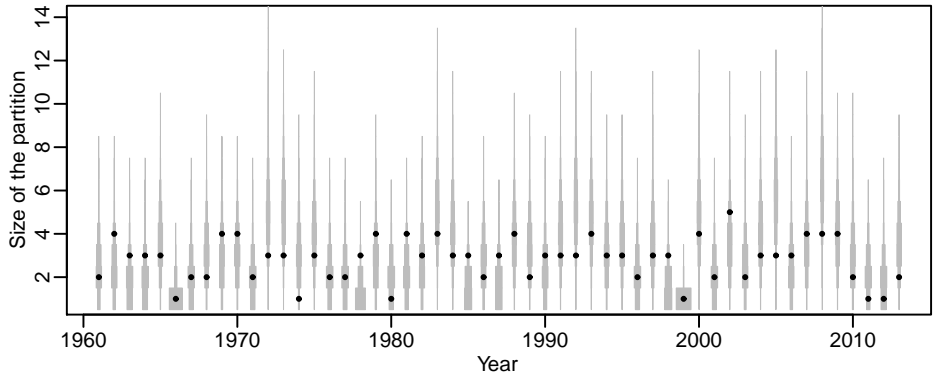}}
\caption{Estimated distributions of the partition sizes for each year. Grey boxes show the relative probability for each size (a wider box corresponds to a higher probability), estimated from the Gibbs sampler. The black dots show the sizes of the partitions obtained using the declustering procedure discussed in~\S\ref{sec:declustering}.}
\label{fig:clust}
\end{figure}

We explain this by the use of the joint model for the minima and the partitions, and the declustering. The first of these requires good agreement of the data with the distribution of the block maxima and the partitions, and this is a strong assumption in practice. Although the Brown--Resnick dependence structure is quite flexible for modeling stationary isotropic dependence structures, the distribution of the underlying partitions can be very different from the partitions obtained through the declustering. In the univariate case, previous studies have shown that  declustering can bias estimators of the marginal parameters if the clusters are poorly-defined, and we can expect similar problems in the multivariate case. In our application, assuming that data separated by more than five days are independent in pairs results in partitions of small sizes each year; see Fig.~\ref{fig:clust}. A similar size for the true partitions for a Brown--Resnick process can only be obtained if the dependence in the process is very strong, and this is incompatible with the empirical extremal coefficients shown in Fig.~\ref{fig:extremalcoef}. Fig.~\ref{fig:clust} also shows the sizes of the random partitions in the MCMC algorithm for model M3; in several cases the size of the partitions from the declustering procedure have low probability for the random partition model. To further compare the partitions from the MCMC algorithm with that from the declustering procedure, we computed the Rand index \citep{Rand:1971}, which is, for each year, the proportion of pairs of locations that are simultaneously either together or separate in both partitions; we then averaged the Rand index over the years. It varies from $0.52$ to $0.63$, indicating that pairwise classification agrees for less than 63\% of the pairs between M2 and M3. The size of the partitions is typically larger in the random partitions, and the relation with the occurrence dates is unclear. For example, in the winter 1973--1974, the winter minima at the 16 stations with no missing data occurred on four consecutive days and the declustering procedure finds a unique partitioning set, whereas just 7\% of the partitions visited in the MCMC run have a unique set and 37\% have four partitioning sets or more.

These results suggest that a different declustering procedure must be used. We varied the number of days used in our declustering scheme, and also considered another scheme that forces the stations that are at least 150~km apart to be in different clusters. The parameter estimates changed with the scheme but the diagnostic plots were never very satisfactory. 

\section{Conclusion}

We have shown how Bayesian inference can be conducted for the Brown--Resnick process, either using the occurrence times of maxima and a declustering procedure, or by including the partitions in an MCMC algorithm.

The partitions add powerful information into the model about the dependence structure. However our application shows that using fixed partitions can badly bias the estimation of the marginal and dependence parameters of a max-stable process. In a second approach we included the partitions in the MCMC algorithm without using information on the occurrence times, and this produced superior results. A more complex approach could use the occurrence times to construct a prior distribution on the space of the partitions, to add more information into the model and improve the efficiency of the estimation, while accounting for the uncertainty on the partitions, but this might be rather complex. \citet{Wadsworth:2015} proposed using second-order terms in the Stephenson--Tawn model to reduce potential bias due to the partitions, but this seems difficult in our case because the number of terms added to the likelihood would be huge and the MCMC algorithm would become even slower.

Our Bayesian model can be generalized in several ways. We focused on a stationary isotropic Brown--Resnick process for simplicity, and it appears to be sufficient for our application. A similar Bayesian hierarchical model can be constructed for the extremal-$t$ process using ideas of \citet{Thibaud.Opitz:2015}, anisotropy would be easily incorporated using the idea of \citet{Engelke.etal:2012}, and so perhaps could the work of \citet{Huser.Genton:2016} on non-stationary dependence models for extremes. However, the generalization of the present method to more complex models and larger datasets may be limited by computational feasibility. Monte Carlo estimation of the likelihood is critical to keep the running times reasonable, and is relatively fast and accurate in low dimensions, but with more sites, more simulations may be needed to obtain reliable inference. 

The model predicts an increase of $0.6^\circ$C per decade in the mean winter minimum temperatures in northern Fennoscandia. Further work to validate the predictions from our model and to quantify the possible effects of global change on boreal forests  would be valuable. The linear trend in time provides sensible short-term predictions, but for longer periods more complex approaches would be necessary, though limited by the availability of reliable data. One possibility would be to link the changes in extreme temperatures to mean temperatures and use predictions from global climate models under different scenarios as inputs to forecasting.

\section*{Acknowledgements}
This work was undertaken while E.T. was at EPFL, Switzerland, with support from the Swiss National Science Foundation. E.T. and D.S.C.'s work at CSU was partially supported by US National Science Foundation Grant DMS-1243102. The temperature data are from the Finnish Meteorological Institute database for Finland, and from the European Climate Assessment \& Dataset database (http://www.ecad.eu) for Sweden and Norway. This work was motivated by discussions with Seppo Neuvonen, who pointed out to us the relevance of winter minimum temperatures to insect outbreaks.

\begin{supplement}[id=suppSM]
  \stitle{Supplementary Material for Bayesian inference for the Brown--Resnick process, with an application to extreme low temperatures}
  \slink[doi]{}
  \sdatatype{.pdf}
  \sdescription{The online supplement provides additional results for the exploratory analysis and the MCMC algorithm used to fit the two Brown--Resnick models, and reports additional diagnostics for the fit of the three models considered in the manuscript.}
\end{supplement}

\appendix



\end{document}